\documentclass[aps,prd,twocolumn,superscriptaddress,altaffilletter,nobibnotes,nofootinbib,10pt,showpacs,showkeys,preprintnumbers]{revtex4-2}

\usepackage{float}
\usepackage{graphicx}
\usepackage{dcolumn}
\usepackage[dvipsnames]{xcolor}
\usepackage{bm}
\usepackage{lineno}
\usepackage{float}
\usepackage{amsmath}
\usepackage{soul}
\RequirePackage{multirow}

\usepackage{tikz,xcolor,hyperref}

\definecolor{lime}{HTML}{A6CE39}
\DeclareRobustCommand{\orcidicon}{
	\begin{tikzpicture}
	\draw[lime, fill=lime] (0,0) 
	circle [radius=0.16] 
	node[white] {{\fontfamily{qag}\selectfont \tiny ID}};
	\draw[white, fill=white] (-0.0625,0.095) 
	circle [radius=0.007];
	\end{tikzpicture}
	\hspace{-2mm}
}
\foreach \x in {A, ..., Z}{\expandafter\xdef\csname orcid\x\endcsname{\noexpand\href{https://orcid.org/\csname orcidauthor\x\endcsname}
			{\noexpand\orcidicon}}
}

\newcommand{\RomanNumeralCaps}[1]
{\MakeUppercase{\romannumeral #1}}

\newcommand{\pte}{$p_{\rm{_T}}$}
\newcommand{\pt}{$p_{\rm{_T}}$~}

\newcommand{\pb}{$PbPb$~}

\usepackage{soul}
\usepackage{float}

\usepackage{soul}

\def\ims{0.45}
\usepackage{subcaption}

\begin{document}
\title{A comparative study of hadron-hadron and heavy-ion collision using the $q$-Weibull distribution function}

\author{Rohit Gupta\orcidA{}}
\affiliation{Kashi Naresh Government Post Graduate College, Gyanpur, Bhadohi 221304, Uttar Pradesh, India}
\author{Satyajit Jena\orcidB{}}%
 \email{sjena@iisermohali.ac.in}
 \affiliation{Indian Institute of Science Education and Research Mohali (IISER), Knowledge city, Sector 81, SAS Nagar, India}

\begin{abstract}

Recent results on multiplicity dependent transverse momentum spectra data in different high multiplicity $pp$ collision have opened a window to search for QGP like medium in hadron-hadron collision. In this work we have performed a comparative study of charged hadron spectra in $pp$, $pPb$ and \pb~collision using the $q$ parameter obtained from the $q$-Weibull distribution function. We observed a disparity in the trend of $q$ parameter in high-\pt range. 
\end{abstract}
\maketitle
\section{Introduction}
The search for a new state of matter formed under extreme conditions of temperature and energy density known as Quark-Gluon Plasma (QGP) has long been a driving force for several heavy-ion collision experiments. Different experiments are colliding heavy ions to search for this novel state. When we collide heavy ions moving at ultrarelativistic speed, a QGP state is formed, but for a very short interval of time, and hence it is not possible to directly probe it. However, the presence of a strongly interacting matter will affect the kinematics of final state particle; hence we rely on the kinematic observables such as energy, transverse momentum, pseudorapidity of final state particles to search for QGP. Indirect signatures such as strangeness enhancement, $J/ \psi$ suppression, jet quenching can signal the presence of QGP. Experiments like RHIC and LHC have already announced QGP formation based on results related to these indirect signatures.  

Small collision system such as $pp$ and $pA$ collision have long been used to establish a baseline for QGP searches in the heavy-ion collision. Comparison with such baseline has provided us with different evidence of QGP signatures in heavy-ion collision, for example, jet quenching \cite{Chatrchyan:2011sx}, collective flow \cite{ALICE:2011ab}, quarkonia suppression and regeneration \cite{Chatrchyan:2012np}.

Heavy-ion collisions has so far been the primary focus in the search for the QGP state, as opposed to pp collisions, because the large number of nucleons in heavy ions allows for the creation of high energy densities and temperatures required for QGP formation. However, in high-energy proton-proton (pp) collisions, Multi Partonic Interactions (MPI) \cite{OrtizVelasquez:2013ofg, Deb:2020ezw} play a crucial role and can lead to high multiplicity events. This, in turn, can create high energy densities over a small volume, potentially forming an ideal scenario for Quark-Gluon Plasma (QGP) creation.

Recent results from high multiplicity $pp$ collision show some effects that are so far considered typical of heavy-ion collision. One such result is the observation of collectivity  \cite{Khachatryan:2016txc} in $pp$ collision. This observation suggests a similarity in the hydrodynamical behaviour of the system created in high multiplicity $pp$ and heavy-ion collision. Another intriguing observation is the enhancement in production of strange hadrons compared to pions \cite{ALICE:2017jyt} as measured by the ALICE experiment. Since the constituents of colliding protons are only up $(u)$ and down $(d)$ quarks, so any enhancement in production of strange hadrons could point towards formation of a strongly interacting medium. These results demonstrates the similarity in the underlying physics mechanism responsible for particle production in high multiplicity $pp$ and heavy-ion collision. However, it's crucial to study whether these phenomena can arise from alternative physical mechanisms without QGP, such as string model that don't explicitly assume deconfinement. Here strings phenomenologically represent the QCD field at large distances. High-multiplicity $pp$ collision events create regions with high string density, potentially leading to the formation of "ropes" through a combination of overlapping strings. This rope formation leads to an increase in the effective string tension, enhancing strange particle production as multiplicity increases \cite{Bierlich:2014xba}.
Further, the color reconnection (CR) model in PYTHIA can fairly explain the flowlike patterns observed in $pp$ collision \cite{OrtizVelasquez:2013ofg, Ortiz:2016kpz}. Hence, before drawing conclusions about the formation of QGP in small collision system, it is essential to explore additional experimental signatures, including jet quenching and suppression of high $p_T$ particles, charmonia and bottomonia suppression etc.

These results on high multiplicity $pp$ collision at different energies have opened up a new avenue in high energy physics to search for QGP state.  With data availability at different multiplicities, it will be interesting to analyse the variation of different thermodynamical properties with multiplicity in $pp$ collision and compare it with the variation shown by data from heavy-ion collision experiment.

From standard statistical mechanics, we know that the study of thermodynamical properties of a statistical system require the distribution of energy of its constituent particles. Transverse momentum (\pte), is the component of the momentum of final state particle in direction transverse to the beam direction. It is related to the energy of particles through transverse mass, $m_T$, as:
\begin{equation}
    E = m_T cosh(y)
\end{equation}
and $m_T = \sqrt{m^2 + p_T^2}$. The \pte-spectra of final state particles produced in high energy collision carry important information about the dynamical and thermal properties of the system created in the collision. Hence proper parameterization of \pte-spectra is necessary to extract the important thermodynamical parameters. Interaction of quarks and gluons is governed by the Quantum Chromodynamics (QCD) theory. However, due to very high coupling strength at low energies, it is difficult to apply perturbative QCD to explain the spectra. Hence, we rely on the phenomenological models to study the transverse momentum spectra. 

Several phenomenological models have been developed, however, most widely used are the statistical thermal models. In high-energy collisions, if the system is assumed to be thermally equilibrated, the Boltzmann-Gibbs statistics provides a suitable framework for understanding the energy distribution \cite{Schnedermann:1993ws, Stodolsky:1995ds}. Adding non-extensivity into the standard Boltzmann statistics gives us the Tsallis statistics \cite{Tsallis:1987eu} which hosts an additional parameter $'q'$ which takes care of the non-extensivity in the system. This parameter effectively scales the system, making it possible to use standard statistical methods for high-energy collisions, where particle numbers are much smaller than the Avogadro number. Given the well-established azimuthal anisotropy in systems produced by high-energy collisions, new models have been developed to incorporate flow properties alongside thermal motion. These include the Blast-Wave (BW)  \cite{Ristea:2013ara, Tang:2008ud} model and the Tsallis Blast-Wave (TBW) \cite{Ristea:2013ara, Tang:2008ud} model. While these models are widely used to study transverse momentum spectra, their applicability is largely limited to the low-\pt region \cite{Gupta:2021efj}. They tend to deviate from experimental data at high-\pte, where hard scattering processes dominate particle production. For the purpose of this analysis we need to study the \pt spectra over varying \pt range starting from low-\pt region upto very high-\pt range and hence we have utilized the $q$-Weibull distribution function \cite{Dash:2018qln}. It is a generalization of Weibull distribution \cite{brown1995derivation} and is shown to nicely explain the \pt spectra as well as multiplicity distribution in high energy collision \cite{Dash:2018qln, Nayak:2018avf, Sharma:2018vsy, Behera:2016hak}. The distribution function provides a good fit to the spectra across a wide range, including high-\pt up to 50 GeV/c \cite{Dash:2018qln}. 

In this work, we have performed a comparative study of transverse momentum spectra $(p_T)$ of final state particles produced in heavy-ion collision and in high multiplicity $pp$ collision. We used the data for $p_T$ spectra of charged hadrons in different multiplicity classes produced in $pp$ collision at $\sqrt{s_{NN}}$ $=$ $5.02$, $7$ and $13$ TeV \cite{Acharya:2019mzb, Acharya:2018orn} along with various centrality \pb collision at $\sqrt{s_{NN}}= 2.76$ and $5.02$ TeV \cite{Abelev:2012hxa, Acharya:2018qsh}. 

In the next section, we will discuss the model used to study the transverse momentum spectra and then we will present the results of our analysis. 

 \begin{figure*}
       \centering
\begin{subfigure}[b]{0.32\textwidth}
\includegraphics[width=\textwidth]{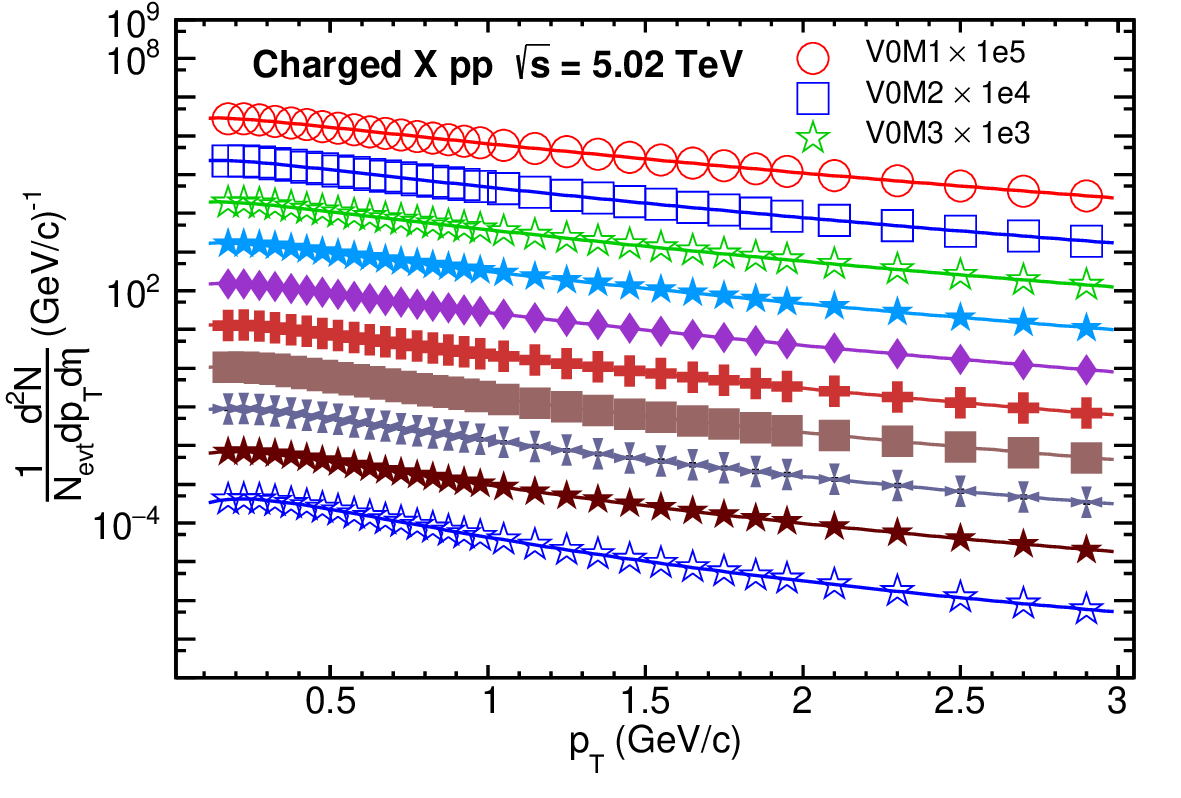}
\caption{$0.15<p_{T}<3$ GeV/c}
\end{subfigure}
\begin{subfigure}[b]{0.32\textwidth}
\includegraphics[width=\textwidth]{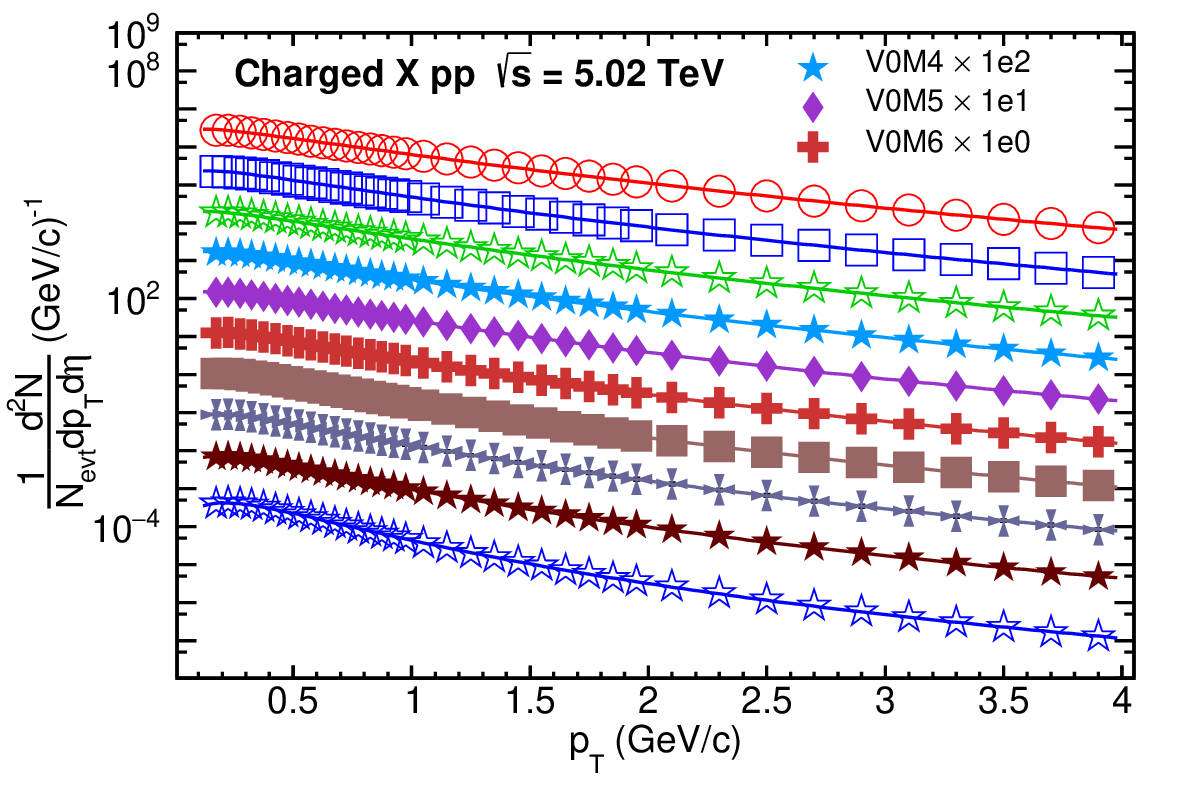}
\caption{$0.15<p_{T}<4$ GeV/c}
\end{subfigure}
\begin{subfigure}[b]{0.32\textwidth}
\includegraphics[width=\textwidth]{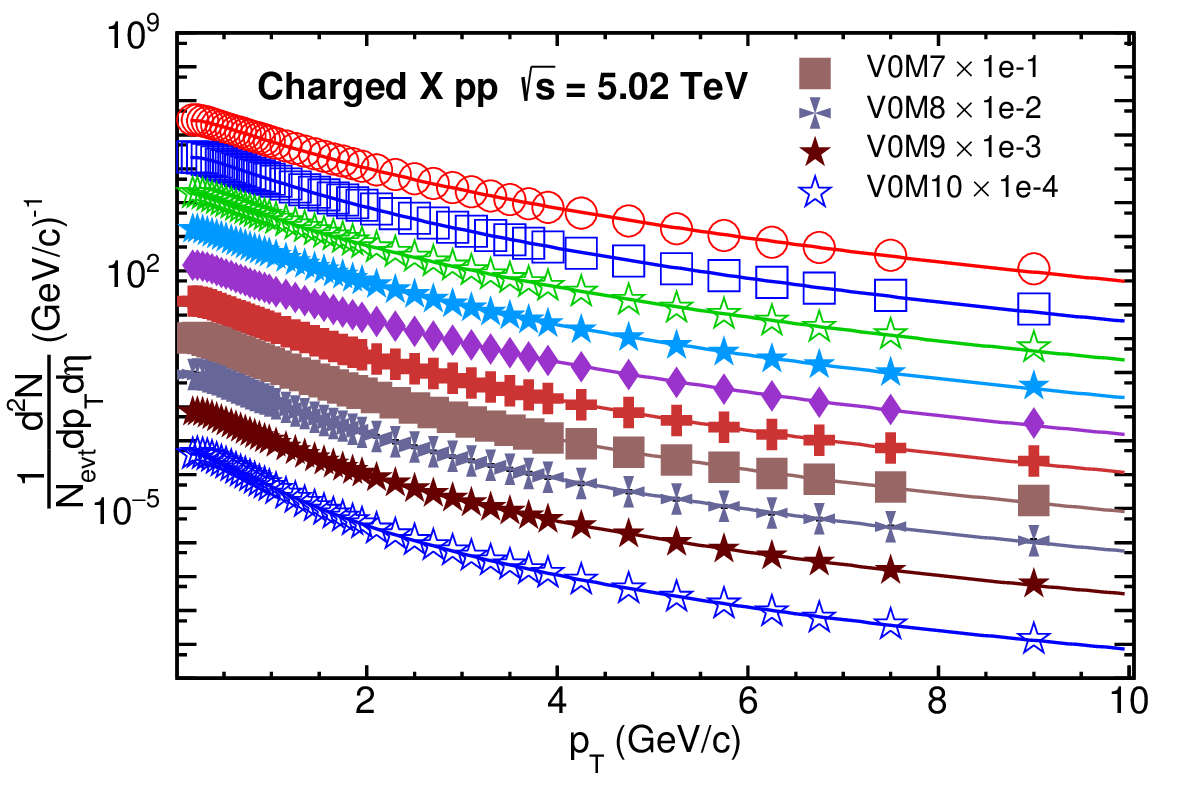}
\caption{$0.15<p_{T}<10$ GeV/c}
\end{subfigure}
\caption{Transverse momentum spectra of charged hadron produced in $pp$ collision at 5.02 TeV \cite{Acharya:2019mzb} fitted with $q$-Weibull distribution function for different $p_T$ range.}
\label{fig:pt_pp_5020}
    \end{figure*}
    
 \begin{figure*}
       \centering
\begin{subfigure}[b]{0.32\textwidth}
\includegraphics[width=\textwidth]{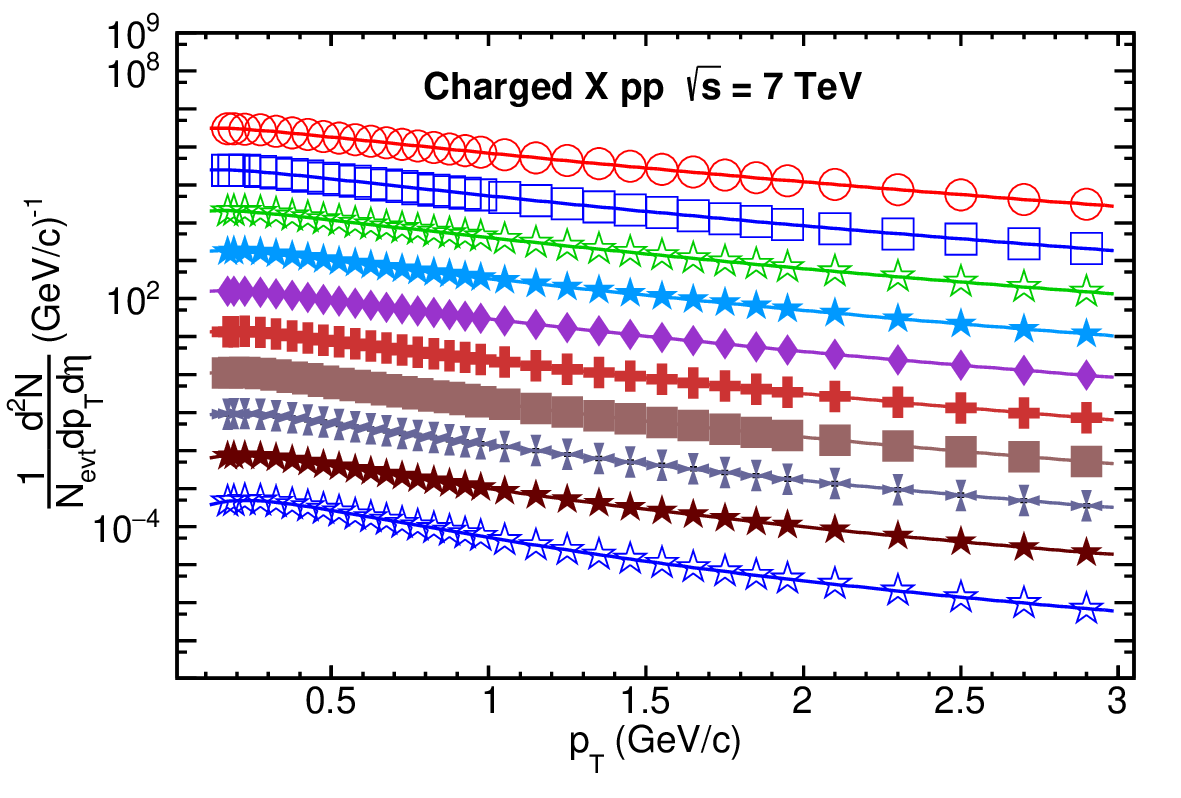}
\caption{$0.16<p_{T}<3$ GeV/c}
\end{subfigure}
\begin{subfigure}[b]{0.32\textwidth}
\includegraphics[width=\textwidth]{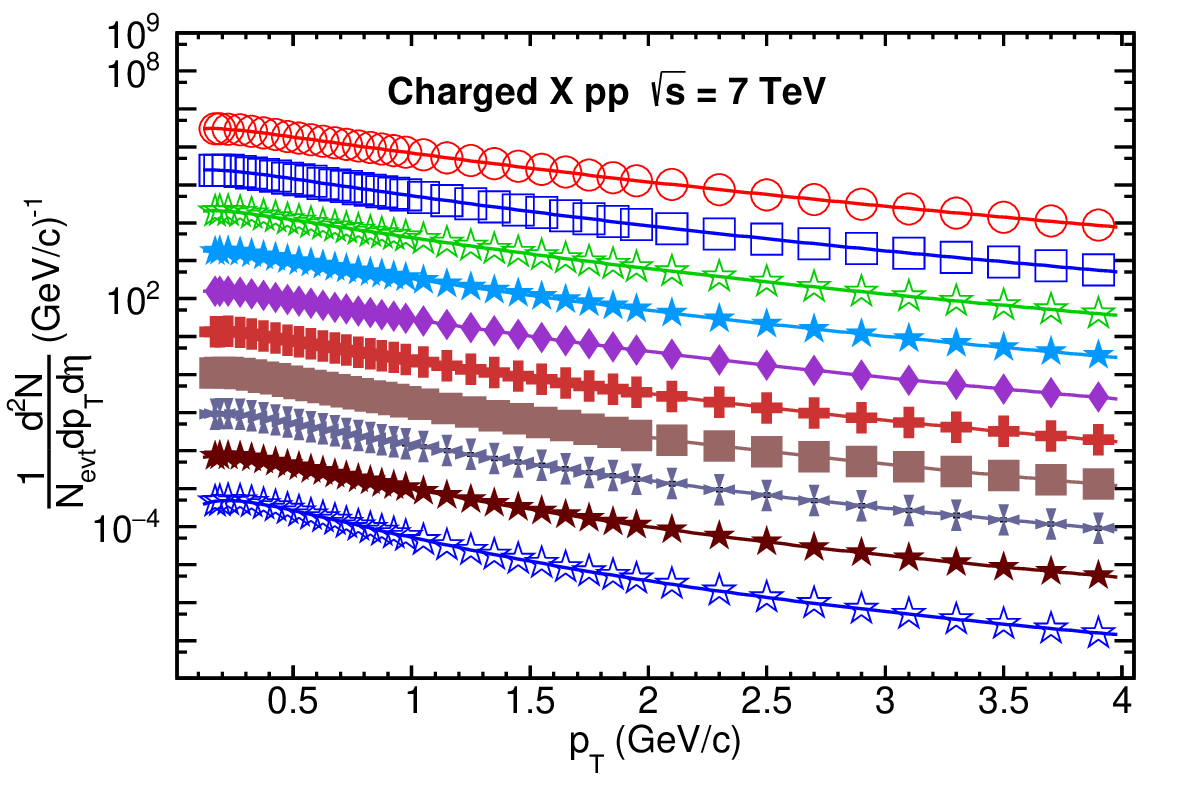}
\caption{$0.16<p_{T}<4$ GeV/c}
\end{subfigure}
\begin{subfigure}[b]{0.32\textwidth}
\includegraphics[width=\textwidth]{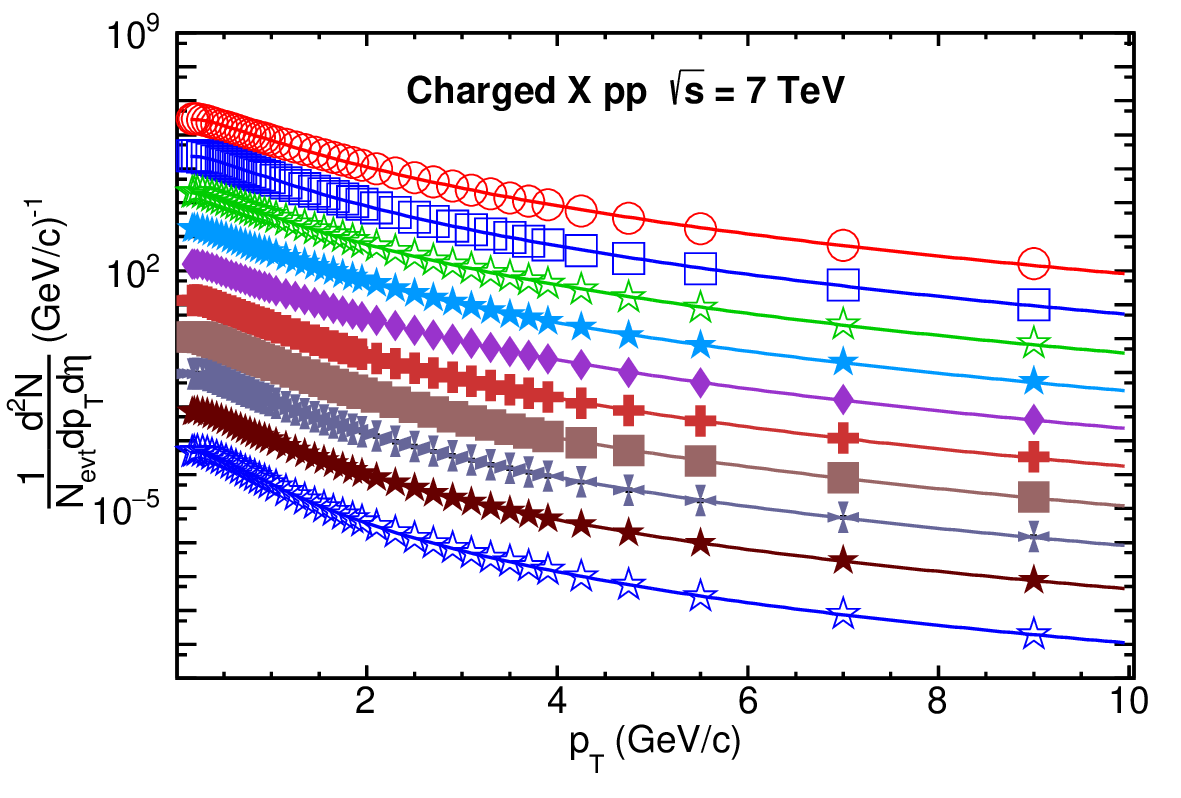}
\caption{$0.16<p_{T}<10$ GeV/c}
\end{subfigure}
\caption{Transverse momentum spectra of charged hadron produced in $pp$ collision at 7 TeV \cite{Acharya:2018orn} fitted with $q$-Weibull distribution function for different $p_T$ range.}
\label{fig:pt_pp_7000}
    \end{figure*}
 \begin{figure*}
       \centering
\begin{subfigure}[b]{0.32\textwidth}
\includegraphics[width=\textwidth]{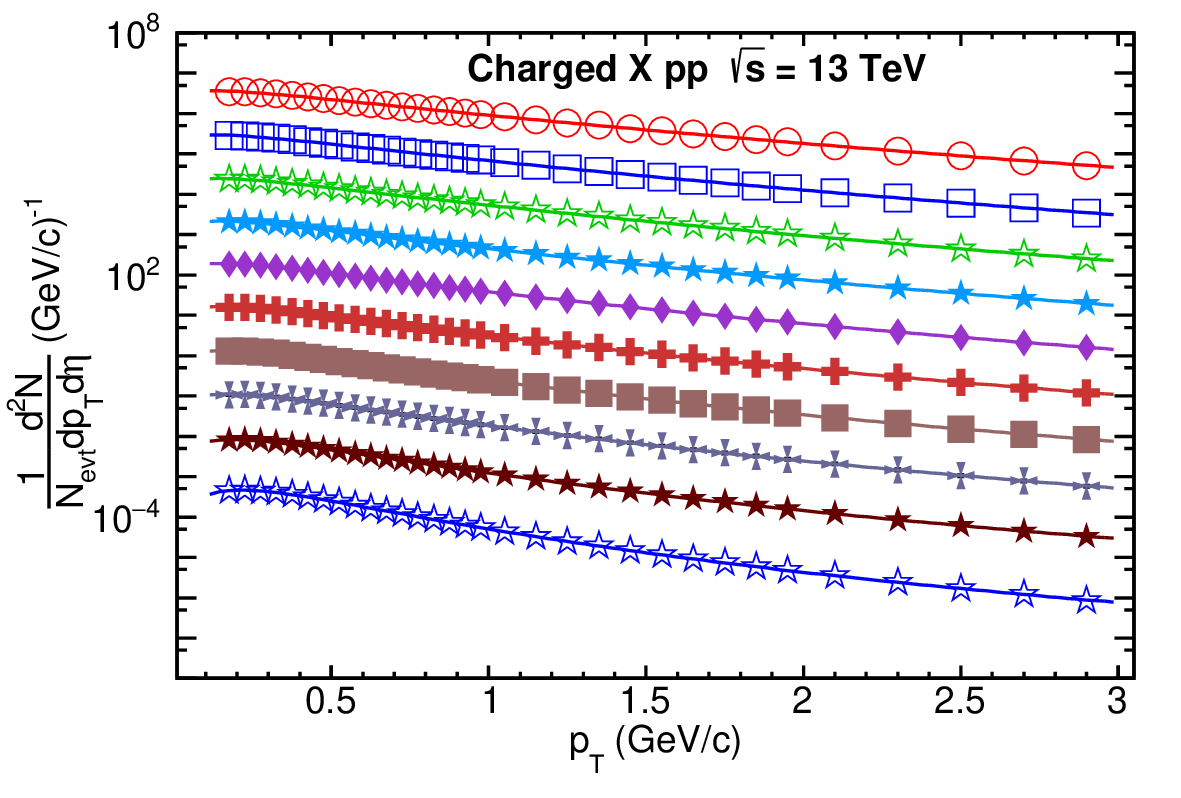}
\caption{$0.15<p_{T}<3$ GeV/c}
\end{subfigure}
\begin{subfigure}[b]{0.32\textwidth}
\includegraphics[width=\textwidth]{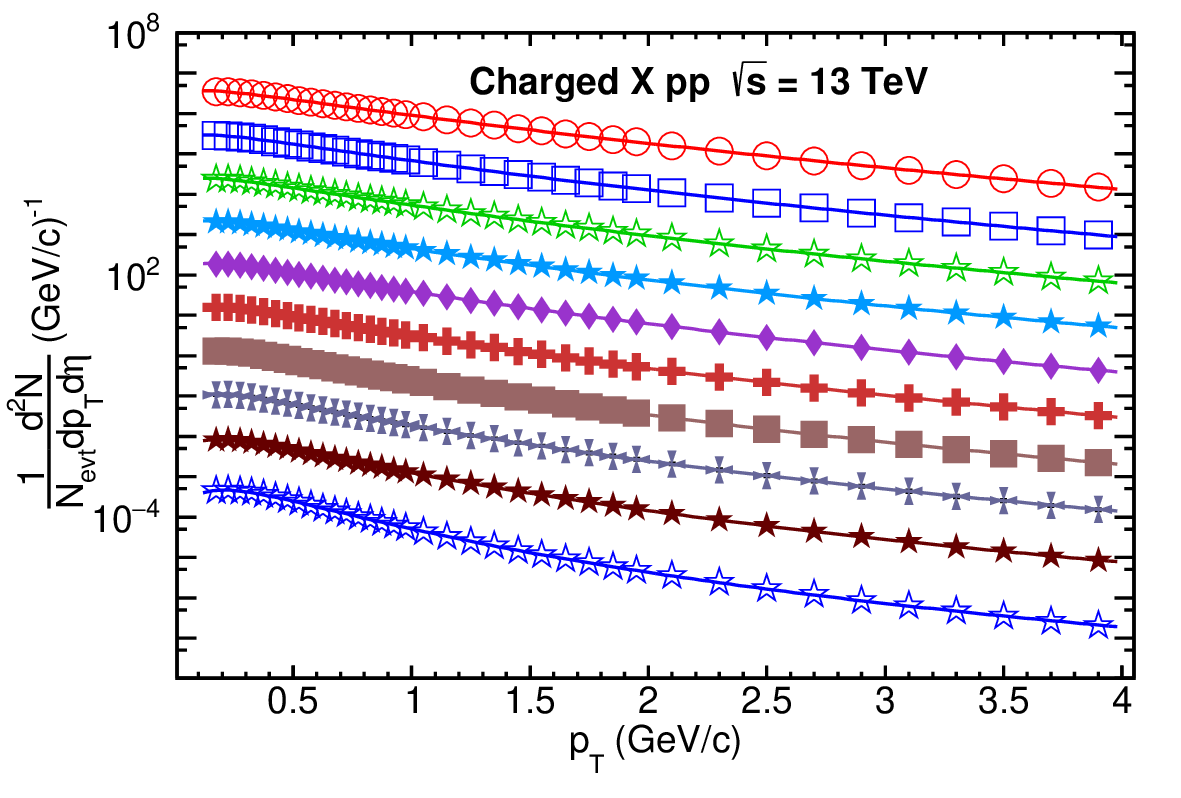}
\caption{$0.15<p_{T}<4$ GeV/c}
\end{subfigure}
\begin{subfigure}[b]{0.32\textwidth}
\includegraphics[width=\textwidth]{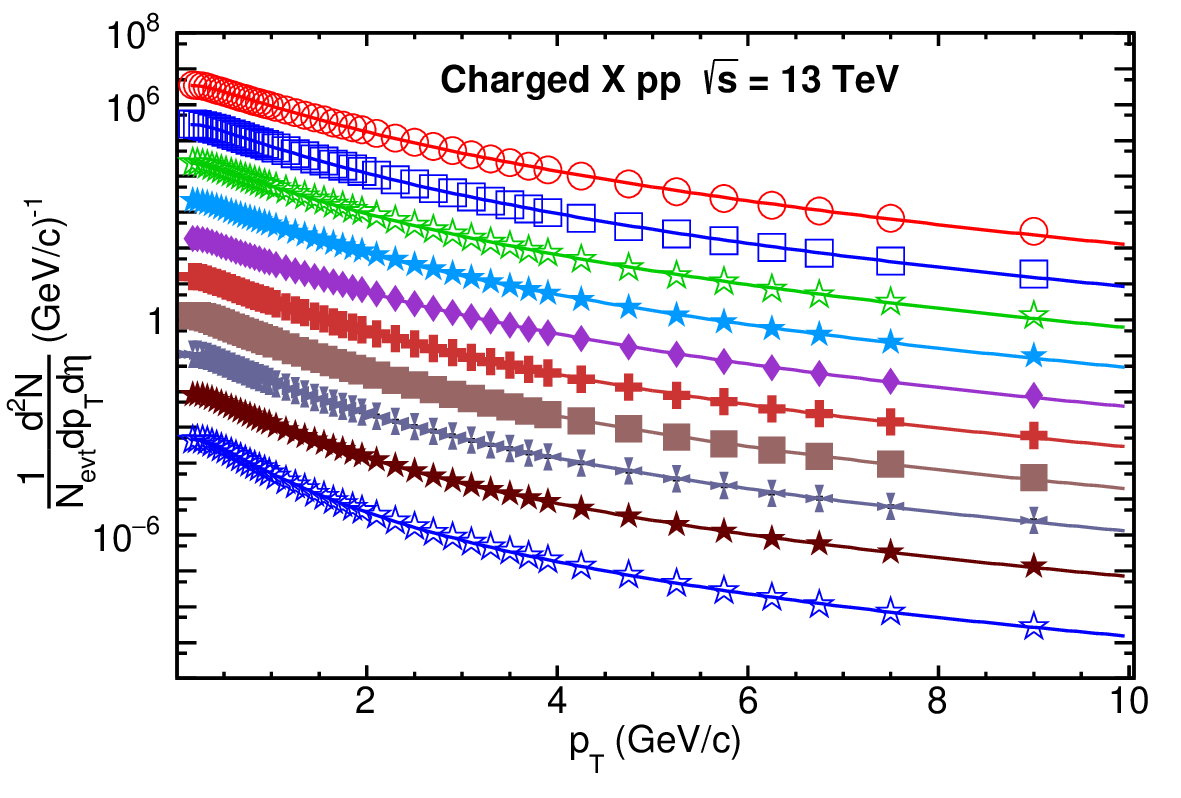}
\caption{$0.15<p_{T}<10$ GeV/c}
\end{subfigure}
\caption{Transverse momentum spectra of charged hadron produced in $pp$ collision at 13 TeV \cite{Acharya:2019mzb} fitted with $q$-Weibull distribution function for different $p_T$ range.}
\label{fig:pt_pp_13000}
    \end{figure*}
     \begin{figure*}
       \centering
\begin{subfigure}[b]{0.32\textwidth}
\includegraphics[width=\textwidth]{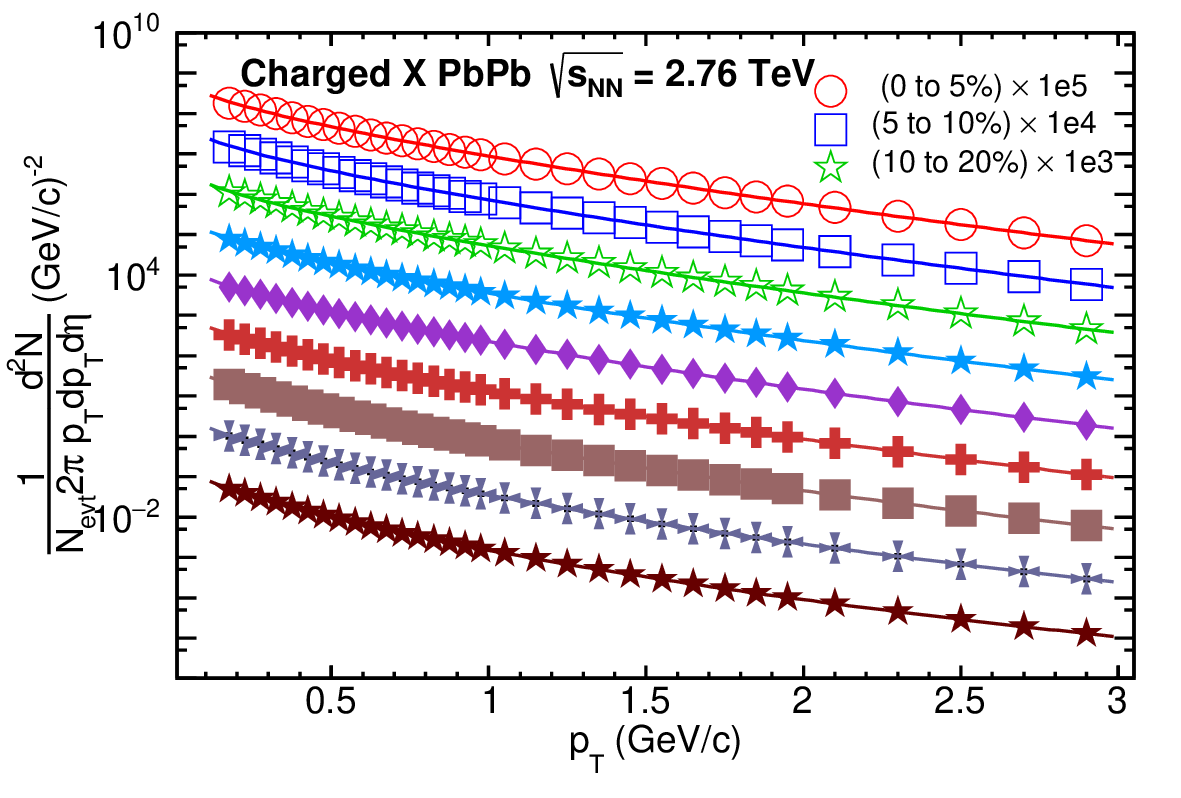}
\caption{$0.15<p_{T}<3$ GeV/c}
\end{subfigure}
\begin{subfigure}[b]{0.32\textwidth}
\includegraphics[width=\textwidth]{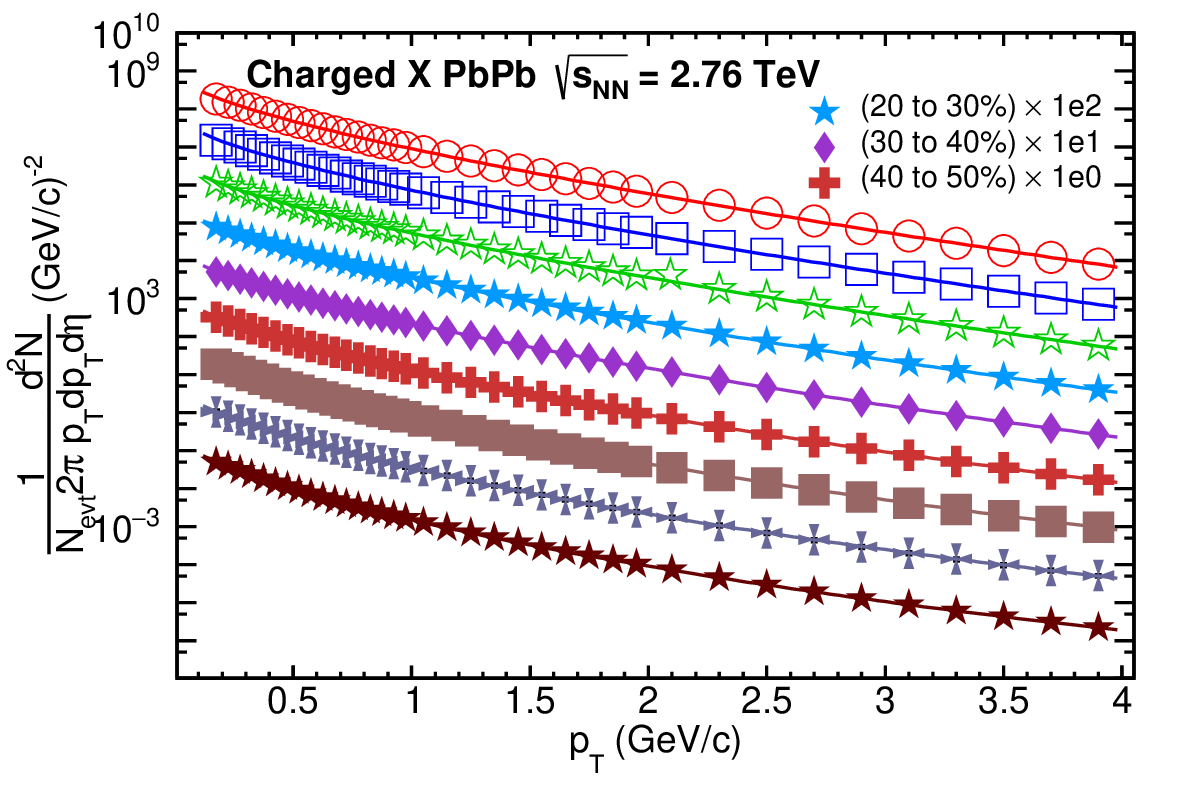}
\caption{$0.15<p_{T}<4$ GeV/c}
\end{subfigure}
\begin{subfigure}[b]{0.32\textwidth}
\includegraphics[width=\textwidth]{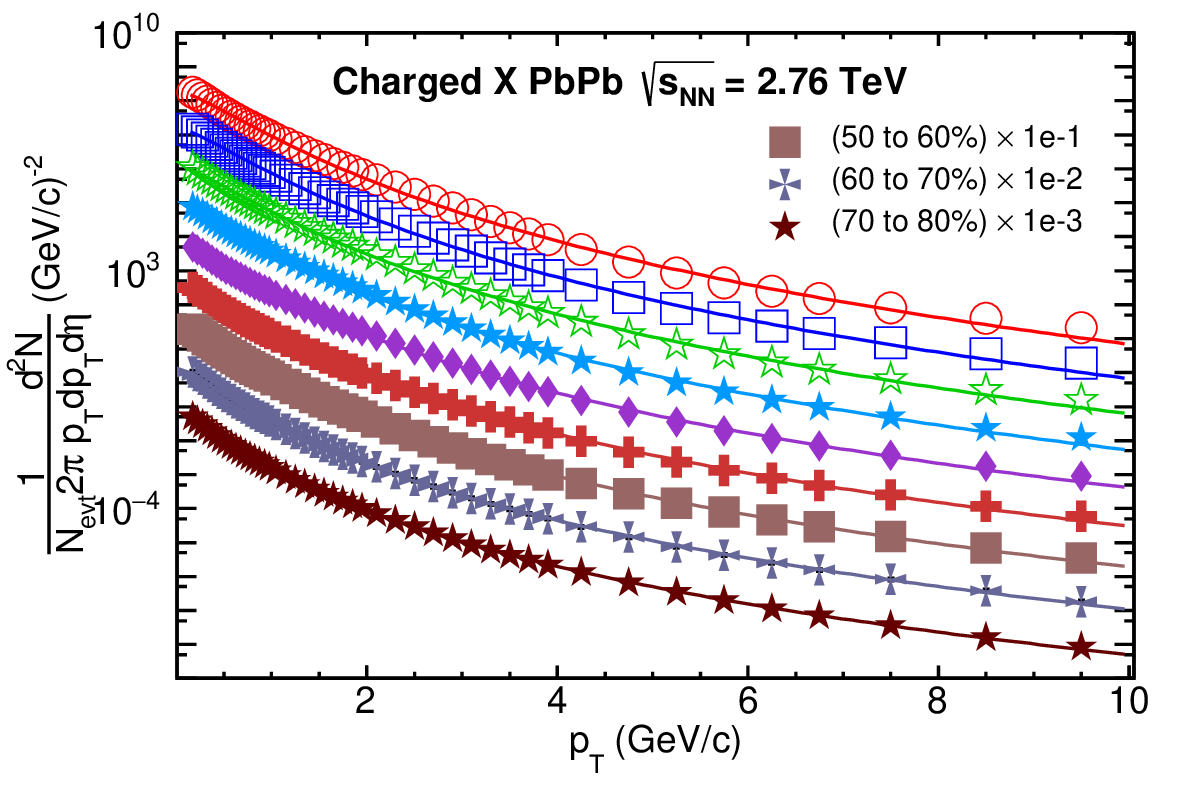}
\caption{$0.15<p_{T}<10$ GeV/c}
\end{subfigure}
\caption{Transverse momentum spectra of charged hadron produced in \pb~collision at 2.76 TeV \cite{Abelev:2012hxa} fitted with $q$-Weibull distribution function for different $p_T$ range.}
\label{fig:pt_pbpb_2760}
    \end{figure*}
\section{$q$-Weibull Distribution}
Weibull distribution is a continuous probability distribution described in 1951 by Swedish mathematician Waloddi Weibull. In Weibull distribution, the probability distribution function is given as:
\[
P(x,\lambda,k)=
\begin{cases}
\frac{k}{\lambda} \left( \frac{x}{\lambda}\right)^{k-1}\; e^{-(x/\lambda)^k} & x \geq 0\\
0 & x<0
\end{cases}
\]
Here $k$ represents the shape parameter and $\lambda$ is the scale parameter of the distribution and $k$ \& $\lambda > 0$.
Weibull distribution has been used previously to describe the the process where the fragmentation and sequential branching drives dynamical evolution of the system \cite{brown1995derivation, brown1989} which makes it a suitable choice to be tested in particle production study. 

Incorporating Tsallis statistics to Weibull distribution gives us $q$-Weibull distribution:
\begin{equation}
P_q(x,q,\lambda,k)=\frac{k}{\lambda} \left( \frac{x}{\lambda}\right)^{k-1}\; e_q^{-(\frac{x}{\lambda})^k}
\label{eqn:31}
\end{equation}
where 
\begin{equation}
e_q^{-(\frac{x}{\lambda})^k} = \left( 1-(1-q)\left( \frac{x}{\lambda}\right)^k\right)^{\left( \frac{1}{1-q} \right)}   
\label{eqn:32}
\end{equation}In the limit $k=1$ and $q\neq 1$ ,Eq.~\ref{eqn:32} reduces to $q$-exponential.

The $q$-Weibull formalism has been used for the first time in Ref.~\cite{Dash:2018qln} to explain the $p_T$ spectra of charged hadrons and it is observed that this model is in good agreement with the $p_T$ spectra over broad range of $p_T$ for both $pp$ as well as heavy-ion collision. The fit parameters extracted from the equation are connected to the underlying characteristics of particle production. The parameter $\lambda$ exhibits an increasing trend from peripheral to central collisions and is connected to the collective expansion velocity of hadrons \cite{Dash:2018qln}. The second parameter in the $q$-Weibull distribution function, $k$, is linked with the onset of hard QCD processes and non-equilibrium scenario. This is also evident from Fig. \ref{fig:k_pt_avg} which shows that the value of $k$ rises with the increasing dominance of hard scattering processes.

One important point to note here is that since $q$-Weibull distribution reduces to exponential distribution only when both $q$ and k reduces to 1, $q$ parameter cannot be directly interpreted as a parameter that quantify deviation from thermal equilibrium. However, $q$ parameters extracted from the Tsallis distribution and the $q$-Weibull distribution show a similar centrality dependence, we can relate the $q$ parameter of the $q$-Weibull distribution to the deviation from thermal equilibrium \cite{Dash:2018qln}.
\section{Results and Discussion}
We begin our analysis by fitting the invariant yield of charged hadrons with $q$-Weibull distribution over different energies for $pp$, $pPb$ as well as \pb~collision. In high energy collision, the particle production in different $p_T$ range is dominated by different process, so the fitting has been performed for different $p_T$ range to study the dependence of fit parameters on the $p_T$ range.  We fitted the $q$-Weibull distribution function with transverse momentum spectra of charged hadrons at different multiplicity of $pp$ collision at $5.02$ TeV, $7$ TeV and $13$ TeV. Table 1, provides the $V0M$ multiplicity class and corresponding multiplicity values for $pp$ collision at three different energies. We also considered the spectra of charged hadrons produced in $2.76$ TeV \pb~collision over different centralities and $pPb$ collision at $5.02$ TeV.  

As is evident from the figures \ref{fig:pt_pp_5020}, \ref{fig:pt_pp_7000}, \ref{fig:pt_pp_13000}, \ref{fig:pt_pbpb_2760}, the predicted values from $q$-Weibull distribution match closely with the actual data. Further, the value of $\chi^2/NDF$ obtained from data-theory comparison also suggest a good agreement between data and theoretical model over broad range of energies and different collision systems.
\begin{figure}[!h]
\centering
  \includegraphics[width=\ims\textwidth]{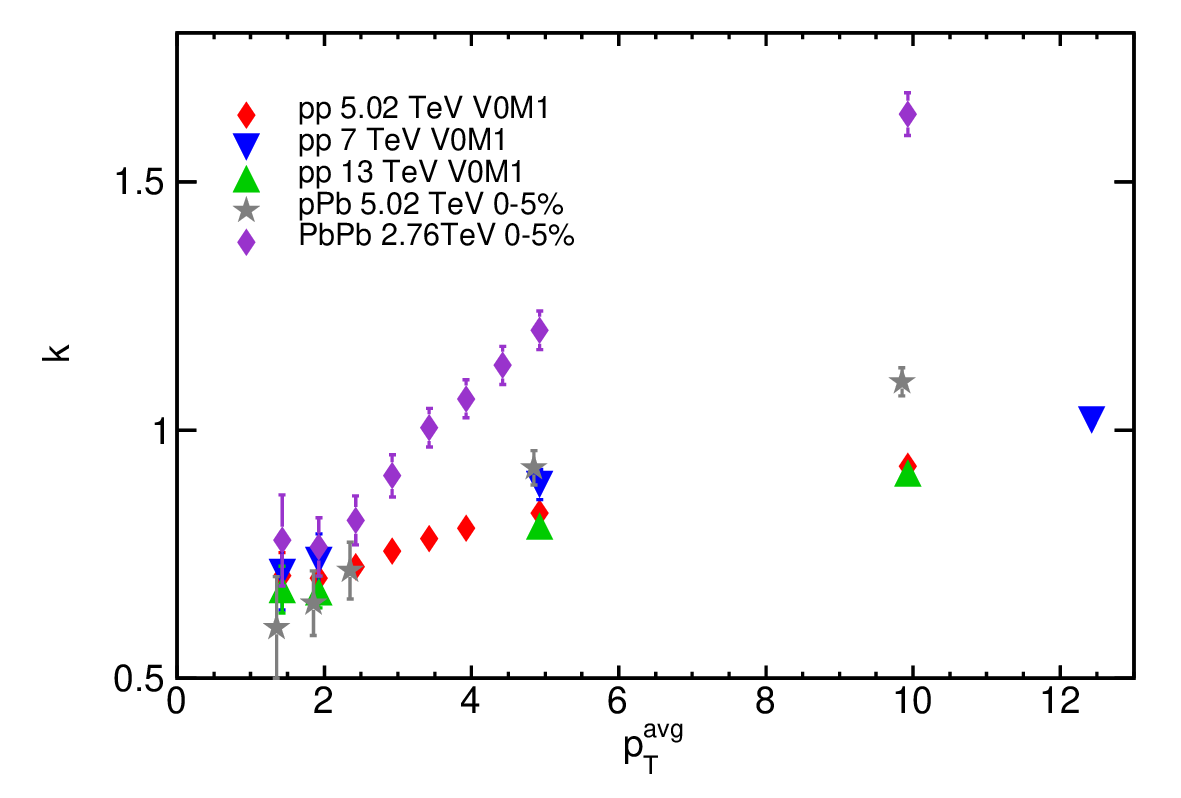}
  \caption{Variation of $k$ parameter with varying $p_T$ range  obtained by fitting most central/highest multiplicity data of charged hadron spectra produced $pp$ \cite{Acharya:2018orn, Acharya:2019mzb}, $pPb$ \cite{Adam:2016dau} and \pb~\cite{Abelev:2012hxa} collision at different energies.}
  
\label{fig:k_pt_avg}
  \end{figure}

   \begin{figure*}
       \centering
\begin{subfigure}[b]{0.45\textwidth}
\includegraphics[width=\textwidth]{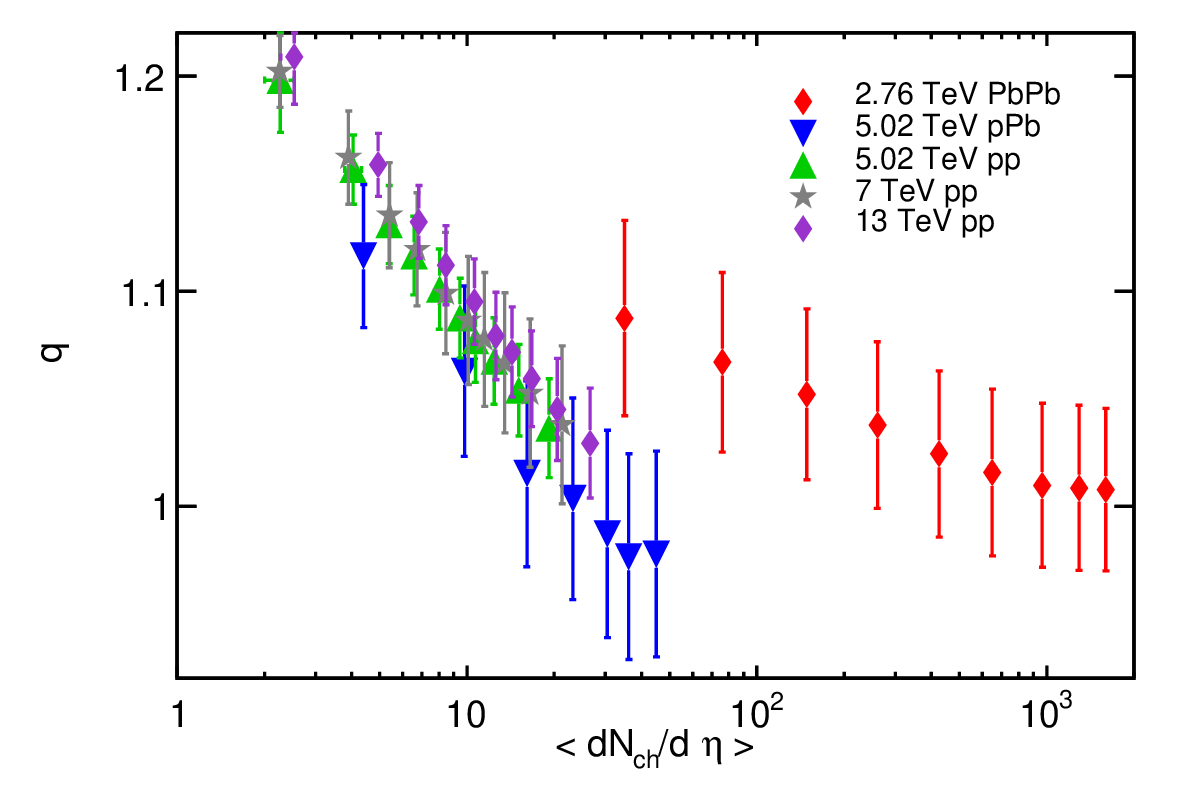}
\caption{$0.15<p_{T}<3$ GeV/c}
\label{fig:q_pt_3}
\end{subfigure}
\begin{subfigure}[b]{0.45\textwidth}
\includegraphics[width=\textwidth]{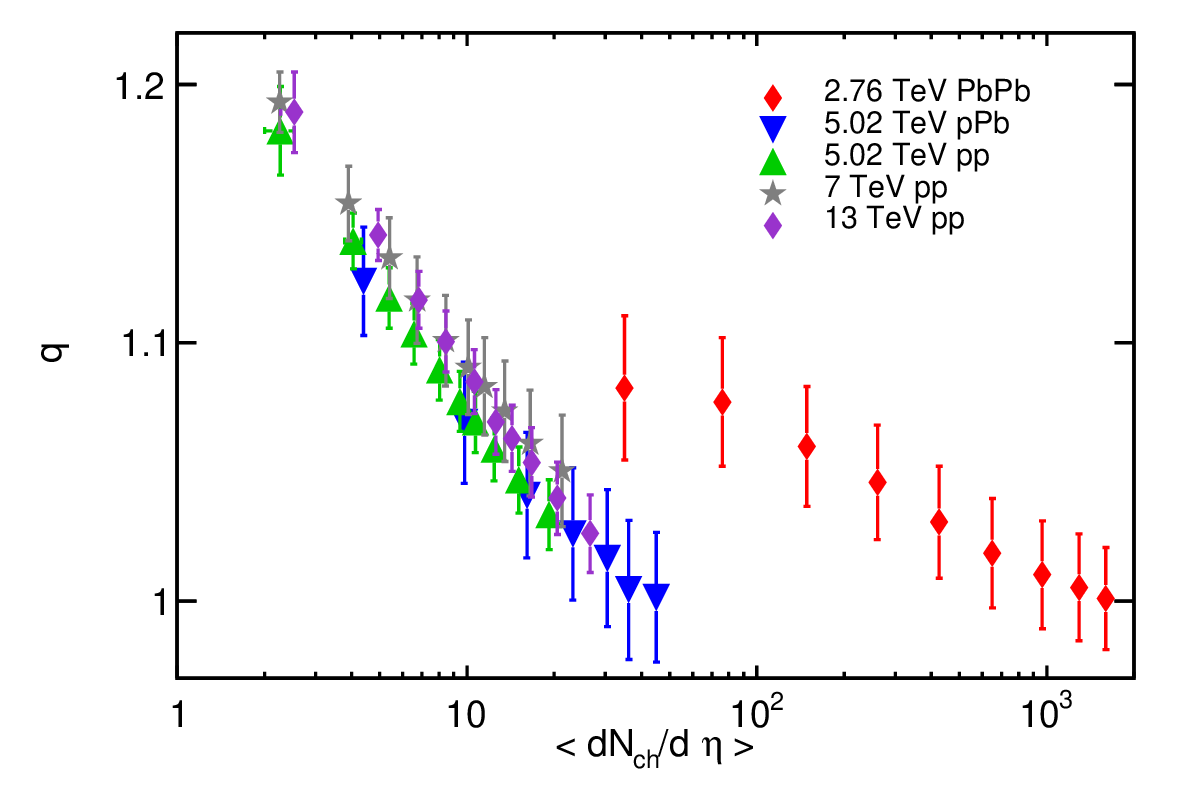}
\caption{$0.15<p_{T}<4$ GeV/c}
\label{fig:q_pt_4}
\end{subfigure}
\begin{subfigure}[b]{0.45\textwidth}
\includegraphics[width=\textwidth]{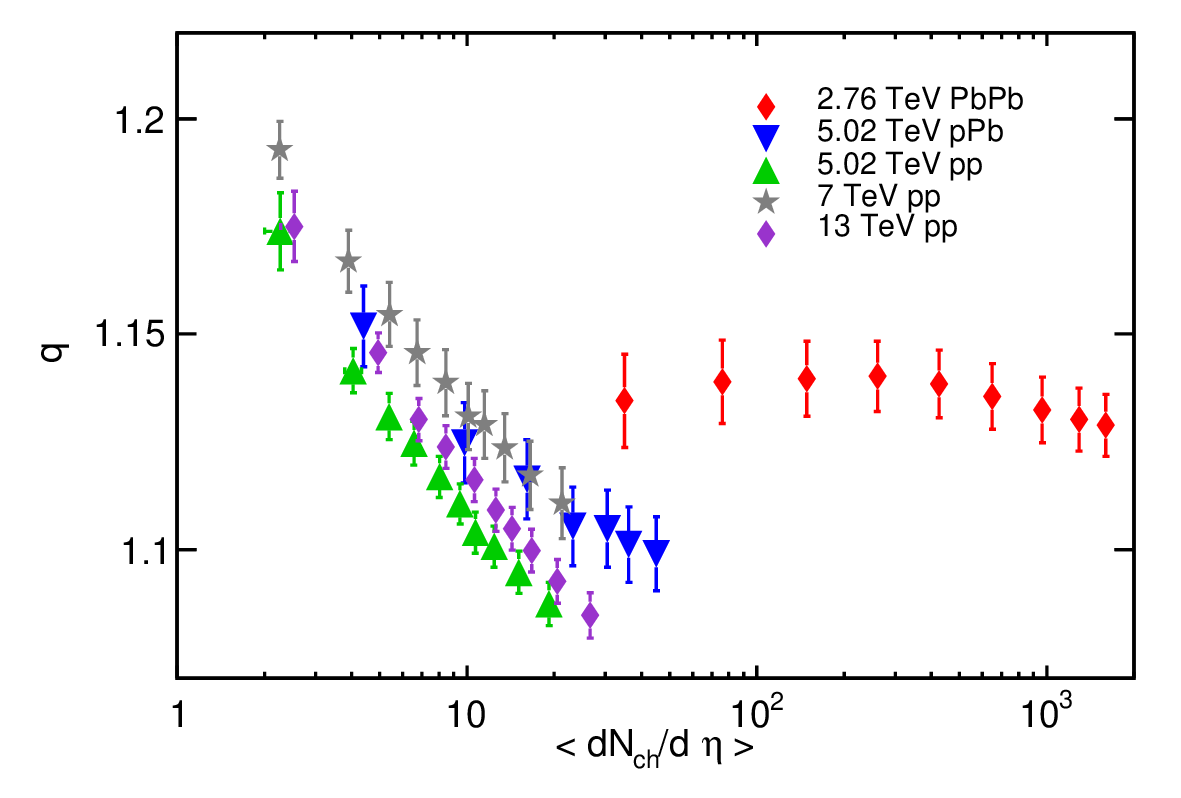}
\caption{$0.15<p_{T}<10$ GeV/c}
\label{fig:q_pt_10}
\end{subfigure}
\begin{subfigure}[b]{0.45\textwidth}
\includegraphics[width=\textwidth]{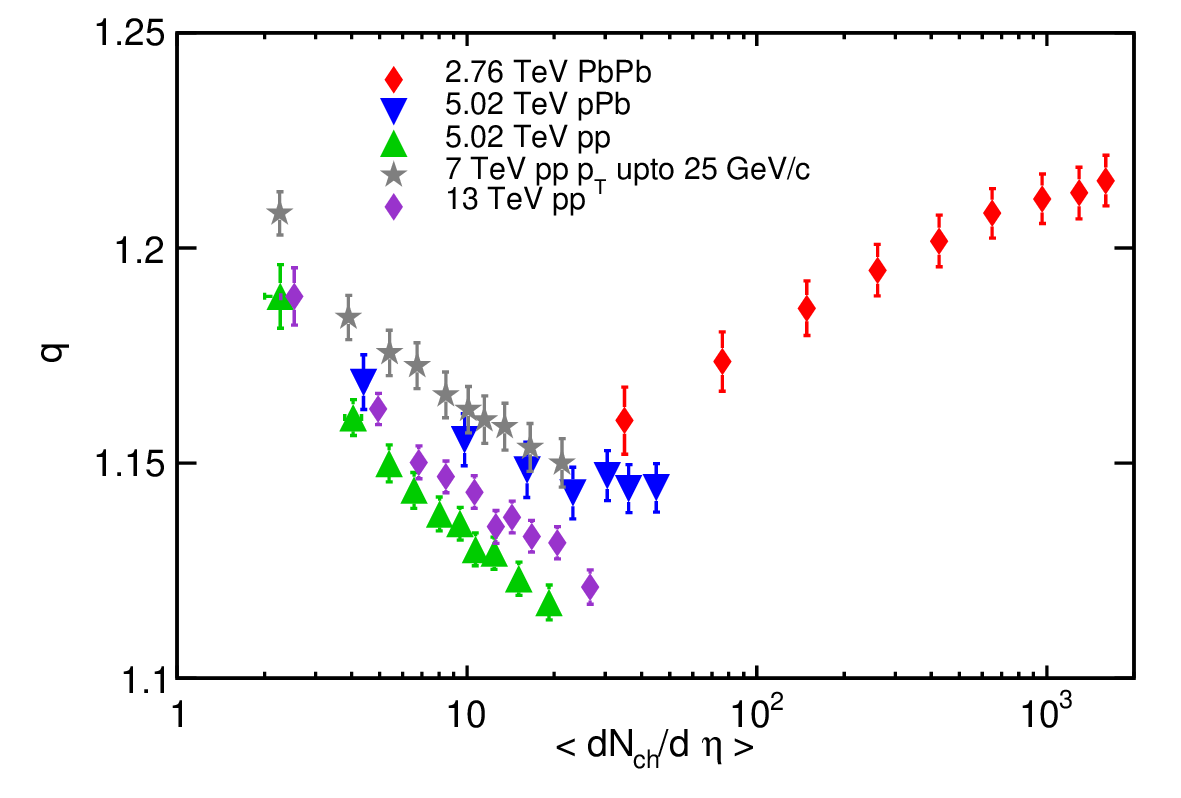}
\caption{$0.15<p_{T}<20$ GeV/c}
\label{fig:q_pt_20}
\end{subfigure}
\caption{Variation of $q$ parameter with multiplicity for different $p_T$ range obtained by fitting charged hadron spectra produced $pp$ \cite{Acharya:2018orn, Acharya:2019mzb}, $pPb$ \cite{Adam:2016dau} and \pb~\cite{Abelev:2012hxa} collision at different energies.}
    \end{figure*}

In Fig. \ref{fig:k_pt_avg} we have plotted the variation of parameter $k$ with the average value of transverse momentum for a given $p_T$ range for most central/highest multiplicity data of $pp$, $pPb$ and heavy-ion collision.  For a $p_T$ range of $p_T^{min}<p_T<p_T^{max}$, the $p_T^{avg}$ is defined as 
 \begin{equation*}
     p_T^{avg} = \frac{p_T^{max} - p_T^{min}}{2}
 \end{equation*} 
 We observe that the value of parameter $k$ increases with increase in $p_T$ range. We know that the high-\pt particles are produced primarily in hard QCD scattering, so, from the trend of $k$ it can be inferred that this parameter is related to the onset of hard scattering process and non-equilibrium conditions as also discussed in Ref. \cite{Dash:2018qln}.
  
We further analyzed the data for different $p_T$ range to observe if there is any significant difference in the pattern of variation of $q$ parameter with multiplicity. In figures \ref{fig:q_pt_3}, \ref{fig:q_pt_4}, \ref{fig:q_pt_10} \& \ref{fig:q_pt_20}, the value of fit parameter $q$ has been plotted with multiplicity for $p_T<3$, $p_T<4$, $p_T<10$ and $p_T<20$ GeV/c respectively. From Fig. \ref{fig:q_pt_3} and \ref{fig:q_pt_4}, we observe that, for heavy-ion collision, $q$ value decreases with increase in centrality; however, Fig. \ref{fig:q_pt_20} show a trend reversal in high-\pt range. For small $p_T$ range more central collision (which correspond to high multiplicity) have smaller $q$ value and $q$ value increase with the increase in centrality. This indicates that the system departs from equilibrium as we increase the centrality. This observation agrees with the hydrodynamical picture and is inline with the results obtained from the Tsallis Blast Wave (TBW) fits \cite{Tang:2008ud}. However, in Fig. \ref{fig:q_pt_20}, we see a trend reversal in high-\pt range. We observe highest $q$ value for most central collision and $q$ value decreases as we move toward peripheral collision. This result has also been reported along with a similar trend for charged hadrons spectra in $Au-Au$ $200$ GeV collision in \cite{Dash:2018qln}. This trend reversal is believed to be because of the dominance of hard scattering processes in particle production mechanism at high-\pt regime \cite{Dash:2018qln}. Hence more central collision will be "harder" and have higher initial state fluctuation, and hence they will deviate more away from equilibrium. This effect makes the spectra more like power law making the $q$ values deviate more from the equilibrium value of $1$.

 \begin{table}
\centering
\caption{Value of multiplicity $\langle dN_{ch}/d\eta \rangle$ for different multiplicity classes of $pp$ collision at three different energies}
\begin{tabular*}{\columnwidth}{@{\extracolsep{\fill}}llll@{}}
\hline
\multicolumn{1}{@{}l}{Multiplicity class} &$5.02$ TeV  & $7$ TeV & $13$ TeV\\
\hline
V0M \RomanNumeralCaps{1}           & $19.2 \pm 0.9$  & $21.3 \pm 0.6$       & $26.6 \pm 1.1$          \\
V0M \RomanNumeralCaps{2}            & $15.1 \pm 0.7$     & $16.5 \pm 0.5$       & $20.5 \pm 0.8$         \\
V0M \RomanNumeralCaps{3}          & $12.4 \pm 0.6$ & $13.5 \pm 0.4$       & $16.7 \pm 0.7$         \\
V0M \RomanNumeralCaps{4}            & $10.7 \pm 0.5$       & $11.5 \pm 0.3$      & $14.3 \pm 0.6$        \\
V0M \RomanNumeralCaps{5}       & $9.47 \pm 0.47$       & $10.1 \pm 0.3$         & $12.6 \pm 0.5$          \\
V0M \RomanNumeralCaps{6}          & $8.04 \pm 0.42$     & $8.45 \pm 0.25$        & $10.6 \pm 0.5$         \\
V0M \RomanNumeralCaps{7}   & $6.56 \pm 0.37$        & $6.72 \pm 0.21$         & $8.46 \pm 0.4$           \\
V0M \RomanNumeralCaps{8}             &$5.39 \pm 0.32$  & $5.4 \pm 0.17$        & $6.82 \pm 0.34$          \\
V0M \RomanNumeralCaps{9}             & $4.05 \pm 0.27$  & $3.9 \pm 0.14$       & $4.94 \pm 0.28$          \\
V0M \RomanNumeralCaps{10}             & $2.27 \pm 0.27$  & $2.26 \pm 0.12$        & $2.54 \pm 0.26$          \\

\hline
\end{tabular*}
\end{table}
    
We performed a similar $p_T$ range dependent analysis for different multiplicity data produced in $5.02$, $7$ and $13$ TeV $pp$ collision along with the same for $pPb$ data at 5.02 TeV. From figures \ref{fig:q_pt_3} and \ref{fig:q_pt_4}, we observed that the trend for low-\pt range is similar to what is observed in heavy-ion collision however we do not see a trend reversal for high-\pt range as is shown in figures \ref{fig:q_pt_10} and \ref{fig:q_pt_20}. 

 In $pp$ collision, even for high-\pt range, we observe the lowest $q$ value at highest multiplicity, and $q$ value increases as we move toward low multiplicity. This indicate that the system is more closer to equilibrium at higher multiplicity. 
 
 The observed difference in the trend of $q$ parameter between high-multiplicity $pp$ collisions and heavy-ion collisions suggests a distinct thermodynamic scenario in these two systems. Furthermore, the influence of hard scattering processes on particle production appears to affect $q$ parameter values \cite{Dash:2018qln, Wong:2014uda, Wong:2015mba}, indicating a connection of parameter $q$ to the underlying production mechanism. Several studies has been performed in RHIC and LHC energies to study the mechanism of particle production in high energy collision \cite{BraunMunzinger:2003pwq, Senger:2004xa, Cleymans:1998fq, Bush:1994qm, Srivastava:2019sck, Mohs:2019iee,Werner:2012xh, Buss:2011mx, Andronic:2017pug}. Also there are different Monte Carlo event generators developed with varying physics input to study various aspect of high energy collisions. PYTHIA \cite{Sjostrand:2007gs} and PHOJET \cite{Engel:1995yda} are the class of event generators primarily used to study $pp$ collisions. One of the difference between these two is that the PHOJET uses the dual parton model to describe non perturbative low-$p_T$ part and PYTHIA uses multiple parton-parton interactions. On the other hand, both uses perturbative QCD to describe hard interaction in high-$p_T$ region. HIJING \cite{Wang:1991hta}, UrQMD \cite{Bass:1998ca}, AMPT \cite{Lin:2004en}, HYDJET \cite{Lokhtin:2008xi}, JETSCAPE \cite{Kauder:2018cdt} etc are the class of event generators used to generate heavy-ion collision events. Despite extensive studies on particle production mechanism at RHIC and LHC energies, the differing trends in the $q$ parameter remain an open question and underscore the need for a thorough theoretical investigation into particle production in both high-multiplicity $pp$ and heavy-ion collisions, spanning low to high $p_T$ regimes.
 
 
 Thermalization is an important criterion for QGP formation. It can be achieved through the mutual interaction between the constituents of fireball provided there is sufficient enough volume, which is the case in heavy-ion collision. However, in the small system such as in $pp$ collision, volume is low compared to the heavy-ion collision. The thermalization like scenario in the small system can be achieved by multiple interactions leading to a phenomenon known as Multi Partonic Interaction (MPI) \cite{OrtizVelasquez:2013ofg, Deb:2020ezw}  given that the interacting particles have sufficient momenta. Further, the number of MPI is larger in higher multiplicity classes of $pp$ collision and it goes down with the decrease in multiplicity \cite{Ortiz:2021peu}. And since MPI also impacts high-\pt particles so higher value of MPI might explain the low $q$ value in highest multiplicity class (Fig. \ref{fig:q_pt_20}) of $pp$ collision data.

\begin{figure}[!h]
\centering
  \includegraphics[width=\ims\textwidth]{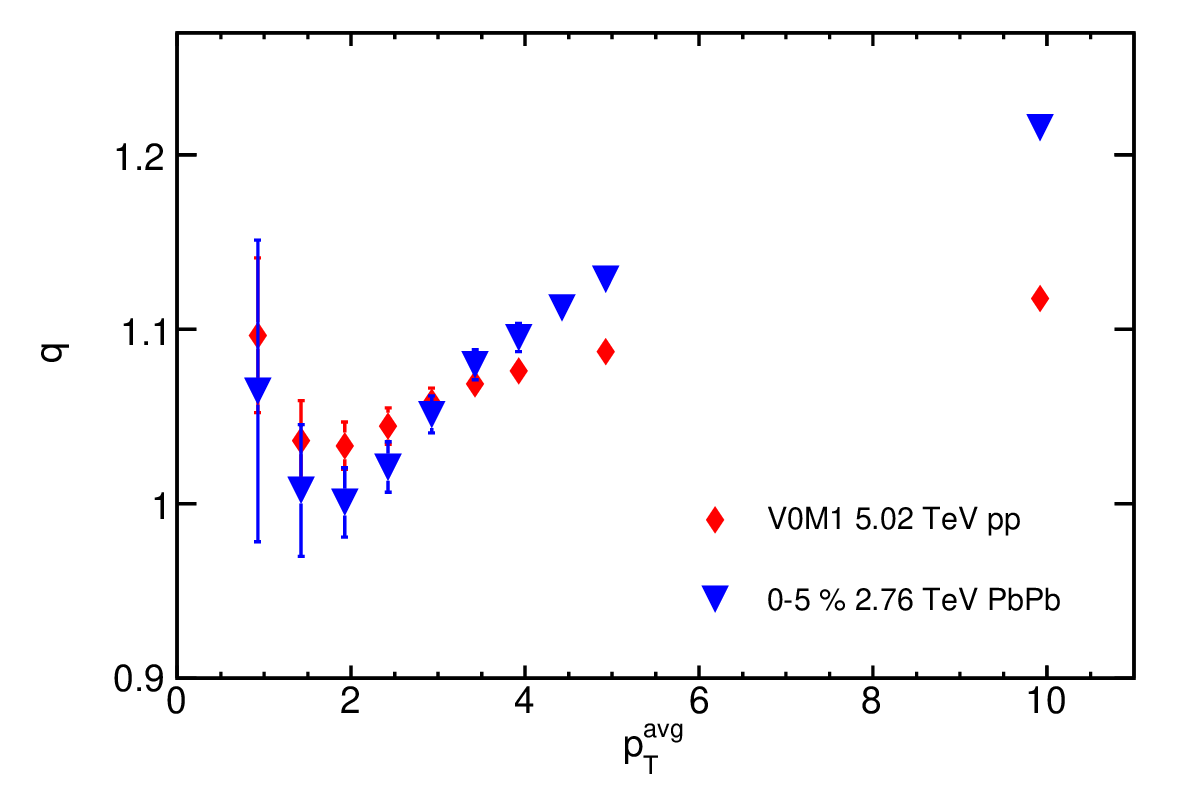}
  \caption{Variation of $q$ parameter with varying $p_T^{avg}$ for highest multiplicity $5.02$ TeV $pp$ collision and most central $2.76$ TeV heavy-ion collision data.}
  
\label{fig:pt_bins}
  \end{figure}

 With an aim to further study the dependence of parameter $q$ on $p_T$ range, we have plotted the value of parameter $q$ versus the average value of transverse momentum for a given $p_T$ range in Fig. \ref{fig:pt_bins}. An intriguing observation that appears from Fig. \ref{fig:pt_bins} is that at a particular multiplicity/centrality, for low-\pt range, the value of $q$ decreases with increase in $p_T$ range. In contrast, it increases with increase in $p_T$ range for high-\pt range both in case of AA and $pp$ collision. For high-\pt range, this observation is obvious as the high-\pt particles come from hard scattering processes leading to an increment in $q$ (more power-law like behaviour). For very low-\pt region, the drop in $q$ value can be explained using the core-corona picture \cite{Kanakubo:2021qcw, Kanakubo:2022ual} developed to get a comprehensive account of high energy collision. In this framework, we can describe the system created during the high energy collision in terms of two components, the core component (equilibrated matter) which is dominant in $p_T\leq 4$ GeV/c and the corona component (non-equilibrated matter) which is dominant in the high-\pt region. However, it is also discussed in the Ref. \cite{Kanakubo:2022ual} that there is some contribution of non-equilibrium component in very low-\pt region because of the fragmentation of strings consisting of hard partons. It is estimated that the corona component reaches upto $20\%$ for $p_T\leq1$ GeV/c. This leads to the non-equilibrium contribution and hence the higher $q$ value in very low-\pt region.
  

\section{Conclusion}
In this paper, we have performed a $p_T$ range dependent study of the $q$ parameter extracted by fitting the transverse momentum spectra of charged hadrons produced in $pp$, $pPb$ and $PbPb$ collision across different energies and multiplicity classes using the $q$-Weibull distribution function. The $p_T$ distribution in heavy-ion collision has particle contribution from two different processes, the particles originating from equilibrated QGP matter in the low-\pt region and the non-equilibrium hard scattering processes in high-\pt region. This leads to a trend reversal in $q$ parameters when we include the higher $p_T$ particles as is evident from the Fig. \ref{fig:q_pt_20}. However, similar analysis for $pp$ collision data does not show any such reversal in the trend of $q$ parameter.  One explanation for this disparity could be linked to the MPI in $pp$ collision MPI tries to thermalise the system and MPI is larger in higher multiplicity classes \cite{Ortiz:2021peu} so we get lower $q$ values even in high-\pt range of $pp$ collision. 

We have also studied the variation of $q$ parameter with $p_T^{avg}$ in  both $pp$ and $PbPb$ collision and observe an opposite trend in very low-\pt and high-\pt region. The increase in high-\pt region can be due to the presence of hard scattering process whereas the decreasing trend in very low-\pt region is explained using the core-corona picture.

In conclusion, we have analyzed the parameter $q$ over different $p_T$ range and observed some disparity in the trend for $pp$ and $PbPb$ collision. There are some physical explanation for the disparity, however, detail theoretical study is necessary to have a better understanding of particle production processes in both $pp$ and heavy ion collision. 
\section{Acknowledgement}
R. Gupta would like to acknowledge that part of this work is carried out using the computing facility in EHEP lab, IISER Mohali.


%
%

%
%

\end{document}